# Study of the Composition and Spectral Characteristics of a HDG-Prism Disperse System (GRISM) by Refractive Index Phase Matching

Chon-Gyu Jo, Chol-Gyu Choe and Song-Jin Im

Department of Physics, **Kim Il Sung** University, Daesong District, Pyongyang, Democratic People's Republic of Korea

**Abstract**

The composition and characteristics of a GRISM gained by refractive index matching between a refractive index modulation type HDG and a prism is investigated, the HDG being built by processing silver halide emulsion with halide vapor. The GRISM has been stable under external influences like humidity or ultraviolet light exposure. The mercury atomic spectrum obtained by a GRISM based on a HDG with a spatial frequency of 600mm$^{-1}$ shows yellow dual lines with the wavelength difference of $\Delta\lambda = 2.1nm$ sufficiently separated.

## 1. Introduction

We have used a holographically obtained phase transmittance type holographic diffraction grating (HDG) as the main disperse unit and integrated it with a wedge prism, using a phase matching method with the consideration of the refractive index of each element, to compose a transmittance type GRISM. It is a straight-line disperse unit useful in a direct vision spectroscope. Then we have commenced a research to examine its spectral characteristics.

Usually a disperse unit obtained by the integration of a HDG and a prism is called a GRISM. [1] Since it has a number of advantages, many studies of its development and use have been published. [2,3] In reference [2] we can find the results of a study on a prism type disperse unit which has a grating pattern covering one of its faces, while reference [3] introduces an application method of a GRISM to a telescopic optical system.

The most common processing method in building a HDG with silver halide emulsion is the method using R-10(KBr) bleach agent. [4,5,6] Usually this method results in a surface relief type HDG. [6,7,8] A phase type HDG obtained by such bleaching method is known to be vulnerable to external environments like ultraviolet exposure. [9,10,11]

Our first step in making a GRISM was to obtain a phase type HDG using $Br_2$ halide steam processing method. [12,13] We have used "лои-2" plate, a silver halide emulsion as the base and processed it with "Kodak D-19" to obtain an amplitude grating. Then we have processed it with $Br_2$ to obtain a refractive-index modulatied phase type HDG. Such a HDG retains micro-scaled relief on its surface which has been existing before the halide steam processing, but it becomes a phase type grating having refractive-index modulation in the emulsion as its main modulation factor. In this case, a phase type HDG with the spatial frequency of 600mm$^{-1}$ has a maximum diffraction efficiency of 60~75%.

During the design of a GRISM as a straight-line disperse unit, we must rationally select geometrical dimensions like the apical angle of the prism.

## 2. Fraunhofer diffraction in a prism-slit disperse system

First, we will discuss the phenomena occuring in an optical system consisting of a slit and a prism, instead of a HDG.

Consider an optical system like figure 1 where a slit of width $a_0$ is attached to a wedge type rectangular prism and let the incident light be a plane light wave. In this condition the light wave entering this system is bound to undergo a continuous phase-change along the slit face. Let the center of

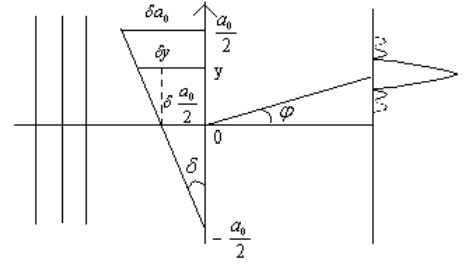

Fig 1. Fraunhofer diffraction in a prism-slit optical system

slit be the origin of coordinates and assume that the apical angle $\delta$ of the prism is not large. Now, at coordinate $y$ the prism's geometric thickness is $\delta \dfrac{a_0}{2} + \delta y$ and its optical thickness is

$\left( \delta \dfrac{a_0}{2} + \delta y \right) n$ where $n$ is the refractive index of the prism's material. The according phase change is

$$\Delta \alpha = kn \left( \delta \frac{a_0}{2} + \delta y \right) + k \left[ \delta a_0 - \left( \delta \frac{a_0}{2} + \delta y \right) \right], \qquad (1)$$

where $k = 2\pi / \lambda$. The first term of Eqs. 1 is the phase change through the prism and the second term is the phase change caused by the optical path $\delta a_0$.

If we consider the Fraunhofer diffraction pattern gained by the slit alone without a prism, since the wave amplitude of a part of the slit of width $dy$ is $A_0 dy / a_0$, the light's complex amplitude gained by the superposition of these waves can be described as the following.

$$A_\varphi = \frac{A_0}{a_0} \int_{-a_0/2}^{a_0/2} e^{-iky \sin \varphi} dy = A_0 \frac{\sin \beta}{\beta}, \dots \qquad (2)$$
$$\beta = k a_0 \sin \varphi / 2$$

If a prism is put before the slit, the amplitude of the light wave diffracted by the slit will undergo a phase change as described in Eqs. 1. Thus it follows

$$A_\varphi = \left( \frac{A_0}{a_0} \right) e^{i[k(n+1)\delta a_0 / 2]} \cdot \int_{-a_0/2}^{a_0/2} e^{-i[k \sin \varphi - k(n-1)\delta]y} dy \qquad \dots (3)$$

and if we put $\beta' = k a_0 \left[ \sin \varphi - (n-1)\delta \right] / 2$ we get the following.

$$A_\varphi = A_0 e^{ik(n+1)\delta a_0 / 2} \frac{\sin \beta'}{\beta'} \dots (4)$$

And it follows

$$|A_\varphi| = |A_0| \left| \frac{\sin \beta'}{\beta'} \right| = |A_0| \frac{\sin\{ka_0[\sin\varphi - (n-1)\delta]/2\}}{ka_0[\sin\varphi - (n-1)\delta]/2} \quad (5)$$

According to the fact that in absence of a prism the amplitude of light in a diffraction pattern is

$$A_\varphi = A_0 \sin\beta / \beta,$$
$$\beta = ka_0 \sin\varphi / 2 \qquad \ldots(6)$$

we can get from Eqs. 5 the conclusion that the primary maximum of the diffraction pattern would deviate as much as

$$\varphi = (n-1)\delta \qquad (7)$$

from the principal axis in a prism-slit optical system.

## 3. Design and composition of a GRISM by refractive index matching

We can use Eqs. 7 to design and compose an optical system where Bragg's diffraction angle for maximum diffraction efficiency is balanced by a prism. In more detail, we must compose the system so that the deviation angle $\varphi$ is balanced by the diffraction angle $\varphi_1$ of the HDG which gives the first order diffraction spectrum, namely the condition $\varphi = \varphi_1$ must be satisfied.

Maximum condition of the diffraction pattern of a refractive index modulation type HDG is satisfied when incident light has Bragg angle as its incident angle.

The Bragg angle can be described as

$$\theta_B = arc \sin\left( \frac{\lambda}{2\Lambda} \right) \qquad (9)$$

where $\lambda$ is the wavelength of light and $\Lambda$ is the spatial period of the HDG.

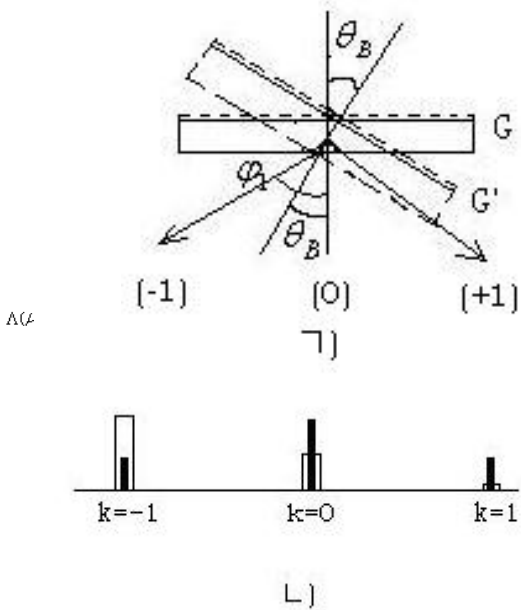

[-1]      [O]      [+1]

ㄱ)

k=-1      k=0      k=1

ㄴ)

Fig 2. light diffraction in HDG

ㄱ) G - grating at vertical incident light

G'- grating at Bragg incident light

ㄴ) black – DE at vertical incident light

white - DE at Bragg incident light

We can derive from Fig 2. that the diffraction angle $\varphi_1$ of the HDG can be expressed as

$$\varphi_1 = 2\theta_B \qquad (10)$$

For the given HDG-prism disperse system it follows from Eqs. (7)~(10) that

$$\varphi_1 = 2\theta_B = 2\arcsin(\frac{\lambda}{2\Lambda}) = (n-1)\delta \quad (11)$$

We derive from Eqs. (11) that the spatial period $\Lambda$ of the HDG and the apical angle $\delta$ of the prism should satisfy the following condition.

$$\Lambda = \frac{\lambda}{2\sin\left[ \frac{(n-1)\delta}{2} \right]} \qquad (12)$$

The relationship between the spatial period of the HDG and the apical angle of the prism is shown in Fig. 3. We can learn from Fig 3. that in case of a HDG with a spatial frequency of $\nu = 300mm^{-1}$ and thereby a spatial period of $\Lambda = 3.33\mu m$, a prism with an apical angle of $\delta = 17°$ is appropriate, while in case of a HDG with a spatial frequency of $\nu = 600mm^{-1}$, and thereby a spatial period of $\Lambda = 1.67\mu m$, a prism with an apical angle of $\delta = 33°$ is appropriate.

The actual matching of a HDG with the refractive index of $n = 1.52$ for its emulsion and a prism with the refractive index of $n = 1.52$ has been conducted by balsam with the same refractive index. (Fig. 4) This method fills the surface relief created while developing the amplitude modulation type HDG and thereafter the HDG acts as a pure refractive index modulated phase type grating. This helps lowering the noise level of the spectrum obtained by it.

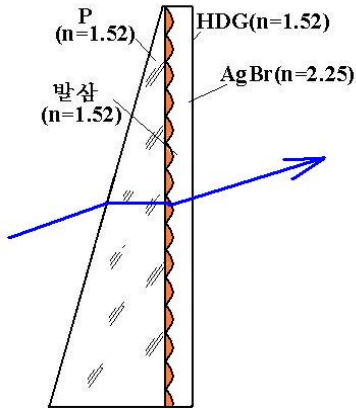

Fig. 4. diagram of the GRISM

(the micro-scale surface relief of the HDG has been magnified.)

processed with Br$_2$

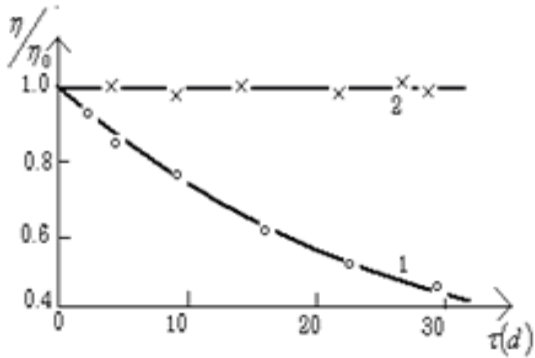

Fig 5. Relative Diffraction Efficiency of the GRISM in an atmosphere with 90% relative humidity

## 4. Characteristics of a GRISM based on a HDG

Normally, a phase type HDG gained by processing a silver halide emulsion with Br$_2$ has little printout effect caused by external ultraviolet exposure but is very vulnerable to a humid atmosphere. [12,13]

A GRISM, the type of which is the combination of the above described HDG and a prism by refractive index matching, has the important trait of being very stable in relation to external conditions.

Fig. 5 shows the stability test result of the GRISM based on Br$_2$ processed HDG objected to humid atmosphere. The figure shows how the diffraction efficiency of the HDG deteriorates to 40% of its initial value during one month in an atmosphere with a relative humidity of 90%, while the GRISM does not undergo any change of diffraction efficiency under the same conditions. Also, under 12 hours of ultraviolet exposure of a high-voltage mercury lamp "дрш-250", a HDG processed with R-10(KBr) undergoes degradation of diffraction efficiency caused by printout effect, whereas the GRISM does not show any signs of such changes. (Fig. 6)

Therefore, the GRISM can be used as a straight-line disperse unit in a direct vision spectroscope.

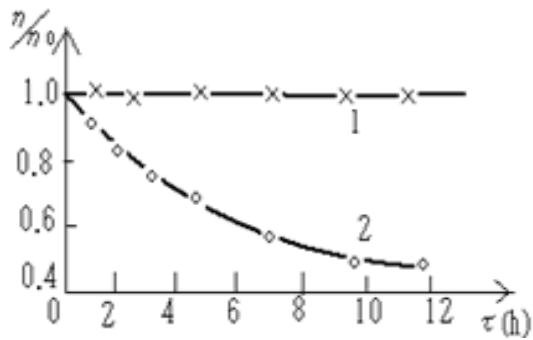

Fig 6. Stability test of the GRISM under ultraviolet light exposure

1- GRISM   2- R-10(KBr) processed HDG

Fig. 7 shows part of the mercury atomic spectrum obtained by the GRISM based on a HDG with the spatial frequency of 600mm$^{-1}$.

As can be seen from Fig. 7 background noise level of the spectrum is quite low and the yellow dual lines $\lambda_1 = 577.0nm$ and $\lambda_2 = 579.1nm$ are sufficiently separated from each other, thus enabling the GRISM to be used as a disperse unit in a direct vision spectroscope.

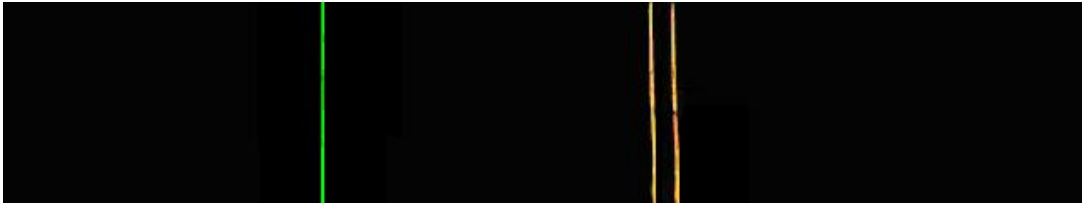

Fig 7. Part of the mercury atomic spectrum obtained by the GRISM based on an HDG with the spatial frequency of 600mm$^{-1}$

## 5. Conclusion

We have, based on a refractive index modulation type HDG gained by processing silver halide emulsion with Br$_2$, composed a GRISM by means of refractive index matching between the HDG and a prism and investigated its characteristics. In case of a HDG with a spatial frequency of $\nu = 600mm^{-1}$ and thereby having a spatial period of $\Lambda = 1.67\mu m$, it is rational to match it with a prism which has an apical angle of $\delta = 33°$. The GRISM is very stable under external influences like humidity or ultraviolet light exposure. The mercury atomic spectrum obtained by a GRISM based on a HDG with a spatial frequency of 600mm$^{-1}$ shows sufficiently separated yellow dual lines with a wavelength difference of $\Delta\lambda = 2.1nm$. The above experimental results show evidence that our GRISM is suitable for efficient usage as a direct-line disperse unit in a direct vision spectroscope.